\documentclass{appolb}
\usepackage{epsfig}
\usepackage{cite}
\usepackage[english]{babel}
\newcommand{\cor}[1]{\left\langle{#1}\right\rangle}

\newcommand{\ve}[2]{\left(\begin{array}{c}\hspace{-7pt}#1\hspace{-7pt}\\
\hspace{-7pt} #2 \hspace{-7pt} \end{array}\right)}
\newcommand{\re}[1]{(\ref{#1})}
\newcommand{\ot}{\otimes}
\newcommand{\ddq}{ { {d \rule{15pt}{0pt}}\over {d\ln{Q^2}} } }
\newcommand{\msbar}{\overline{MS}}

\newcommand{\eq}{\begin{equation}}
\newcommand{\eqx}{\end{equation}}
\newcommand{\eqn}{\begin{eqnarray}}
\newcommand{\eqnx}{\end{eqnarray}}
\newcommand{\dt}{\Delta}

\newcommand{\sd}{\displaystyle}
\newcommand{\nin}{\noindent}
\newcommand{\as}{\widetilde{\alpha_s}(Q^2)}

\newcommand{\dl}{$ln^2(1/x)\,$}

\newcommand{\gp}{g_1^{\gamma}}
\begin{document}
\nin
\begin{center}
{\Large \bf Polarized deep inelastic scattering at low Bjorken $x$ and
resummation of logarithmic corrections, $ln^2(1/x)$.}\\
\vspace{10mm}

{\Large Beata~Ziaja $^{\dag}$}
\footnote{e-mail: ziaja@unix.tsl.uu.se}\\

{\footnotesize

\vspace{3mm}
           $^{\dag}$ \it Department of Theoretical Physics,
	               Institute of Nuclear Physics,\\
                      \it Radzikowskiego 152, 31-342 Cracow, Poland\\
}
\end{center}

\vspace{5mm}
%
%
\nin
\centerline{\it Dedicated to Professor Jan Kwieci\'nski in honour of his 65th birthday}\\ \\
{\bf Abstract:}\\
{\footnotesize
For an accurate description of the polarized deep inelastic scattering 
at low $x$ including the logarithmic corrections, $\ln^2(1/x)$, is required. 
These corrections resummed strongly influence the behaviour of the
spin structure functions and their moments. Results of the
work of J. Kwieci\'nski and myself on this problem are reviewed.}

\section{Introduction}

The problem of identifying the spin components of the polarized nucleon
has been attracting the attention of the high-energy community since
many years \cite{reya,jet3,vetterli,ioffe}.
The data obtained in 1988 by the EMC collaboration \cite{pemc} showed
that the total participation of quarks in the proton spin was very small.
This contradicted theoretical predictions, obtained with the well-founded
Ellis-Jaffe sum rule \cite{vetterli,ioffe}. That sum rule connected the moments of quark
distributions to the nucleon axial coupling constants. Following that rule,
the quarks should participate in about three-fifth of the total
nucleon spin with the parton quark model. The discrepancy between the
theoretical expectations and the experimental data for the polarized proton has
been often referred as the "puzzle of the proton spin" \cite{jet3}.

Since 1988 several observations contributing to this problem have been made.
New experimental data were collected. They showed that neglecting
the participation of the strange quarks at the Ellis-Jaffe sum rule 
was not correct
\cite{vetterli,jet3}.

A significant progress in the theoretical analysis of the nucleon spin 
structure function has been also made. A derivation of $g_1$ in the subleading
approximation of the quantum chromodynamics (NLO DGLAP) showed a difficulty
with the unique defining the quark component of the polarized nucleon.
In the standard $\msbar$  factorization scheme
at the NLO QCD \cite{vann1,vogel} the first moment of the singlet parton
distribution, $\sd \dt \Sigma(Q^2)$, is not conserved. The singlet moments
calculated at different  factorization schemes differ by a term,
$\sd \as \int_0^1 dx \dt g(x,Q^2) \sim O(\alpha_S^0)$, which can be large even
at $Q^2\rightarrow\infty$. This is a direct consequence of the axial anomaly
\cite{axial,leader2,jet3}.
Therefore the differences between the quark moments at different schemes can
be large. This makes the physical interpretation of these moments
difficult.

It is also not clear how the orbital angular momentum, $L$, participates in the
total nucleon spin \cite{jet3}. Again, there is a problem with the unique
definition of this momentum. The total angular momenta of quarks and gluons,
$J_q$ and $J_g$, are well defined and gauge invariant. However, separation
of their spin and orbital components is not unique. It depends on the gauge
transformation used.

New experiments emerge, and they will possibly help explaining the puzzle of the
nucleon spin. The data from the region of low values of Bjorken $x$,
$x<10^{-3}$, will be of special importance. Theoretical predictions show that
at low $x$ the structure function, $g_1(x,Q^2)$, is influenced by large
logarithmic corrections, $ln^2(1/x)$, \cite{BARTNS,BARTS}. As a consequence,
large contributions to the moments of the structure functions from this region
are expected \cite{BARTNS,BARTS,blum9603,blum9606,kiyo,BBJK,kz}.
Including the region of low $x$ into the experimental analysis
will improve the estimation of the parton contributions to the nucleon spin.
This may lead to a better understanding the spin components of the nucleon.

In Refs. \cite{kz,foton,fjet1,fjet2} we investigated how those logarithmic
corrections, $ln^2(1/x)$, influenced the spin structure function, $g_1$, at
low values of Bjorken $x$. We formulated the evolution equations for the
unintegrated parton distributions which included the complete resummation of
the double logarithmic contributions, $\ln^2(1/x)$. Afterwards, those
equations were completed with the standard LO DGLAP evolution terms, in order
to obtain the proper behaviour of the parton distributions at the moderate
and large values of $x$.

The equations obtained were applied to the following observables and
processes: (i) to the nucleon structure function, $g_1$, in the polarized
deep inelastic scattering, (ii) to the structure function of the polarized
photon, $g_1^{\gamma}$, in the scattering of a lepton on a polarized photon,
and (iii) to the differential structure function, 
$x_J\,\partial^2g_1/\partial x_J\,\partial k_J^2$,
in the polarized deep inelastic scattering accompanied by a forward jet. Case
(iii) was proposed to be a test process for the presence and the magnitude
of the $\ln^2(1/x)$ contributions. For each process the consequences of
including the logarithmic corrections were studied in a detail.

Finally, some predictions for the observables, the asymmetry and
the cross sections, in the processes (i)-(iii) were obtained.
They were important to planned experiments with the polarized HERA and linear
colliders, which would probe the region of low values of Bjorken $x$.

\section{Nucleon spin structure function, $g_1$}


The total cross section for the polarized deep inelastic scattering is expressed
by a contraction of two tensors: the leptonic tensor, $L_{\mu\nu}$ and the
hadronic tensor, $W^{\mu\nu}$, $\sigma \sim L_{\mu\nu} W^{\mu\nu}$
\cite{reya}. The hadronic
tensor can be expanded in terms of the Lorentz invariants, and the expansion
coefficients define measurable structure functions \cite{reya,ioffe}.
The spin structure function,
$g_1$, is an expansion coefficient at the asymmetric spin term,
$\epsilon^{\mu\nu\rho\sigma} q_{\rho}S_{\sigma}$, where $q$ denotes the
momentum transfer, and $S$ is a polarization vector of the nucleon.

At low values of Bjorken $x$ the asymptote of $g_1$ predicted by the Regge model
shows a weak scaling with $x$ \cite{regg1,regg2}:
\eq
g_1^{i}(x,Q^2) = \gamma_i(Q^2)x^{-\alpha_{i}(0)},
\label{rg1}
\eqx
where $g_1^{i}(x,Q^2)$ is either a singlet ($i=S$) or a non-singlet
($i=NS$) combination of the nucleon structure functions, $g_1^{p}$ and
$g_1^{n}$. It is expected that $\alpha_{S,NS}(0) \le 0$ and that
$\alpha_{S}(0) \approx \alpha_{NS}(0)$ i.e. the singlet spin
structure function with the Regge model is expected to show similar 
behaviour at low $x$ as the non-singlet one \cite{kz}.

The asymptotic form  of $g_1$ obtained with the standard DGLAP evolution in QCD
is more singular than this obtained with the Regge model \cite{BARTNS,BARTS}.

However, these predictions do not involve the contribution of the logarithms,
$\ln(1/x)$, which is significant at low $x$. In the unpolarized DIS the
resummation of the logarithms, $\ln(1/x)$, was performed by the BFKL equation
\cite{bfkl1,bfkl2}. The analysis of the polarized DIS showed that at low $x$
the structure function of the polarized nucleon, $g_1(x,Q^2)$, was influenced by
large double logarithmic corrections, $ln^2(1/x)$, \cite{BARTNS,BARTS}.
Those corrections originated from the ladder diagrams and the nonladder
("bremsstrahlung") diagrams with the requirement of ordering the ratios,
$k_n^2/x_n$, for the exchanged partons \cite{BARTNS,BARTS,QCD, QCD1}. The momentum,
$k_n$, was the transverse momentum, and the fraction, $x_n$, was the
longitudinal momentum fraction of the $n$th parton exchanged.
Ladder diagrams corresponded to the diagrams describing the forward 
scattering of a photon on a nucleon with a ladder of partons radiatively 
generated \cite{BARTNS,BARTS,BBJK}.
Nonladder diagrams were the ladder diagrams of the forward $\gamma N$ 
scattering with some extra parton propagators attached. 
They were resummed with the infrared evolution equation
\cite{QCD,QCD1}.

Those logarithmic corrections were studied in a detail in Refs. \cite{kz,BBJK}.
Recursive equations for the resummation of the double logarithmic contributions
were there formulated, both for the singlet and the nonsinglet component of
$g_1$. Those equations transformed to the moment space gave proper anomalous
dimensions as derived from the infrared evolution equations
\cite{QCD,QCD1,kz}.

The resummation of \dl was performed for the unintegrated parton
distributions, $f_i$ ($i=S,NS,g$). Therefore
the ordering requirement, $k_n^2/x_n < k_{n+1}^2/x_{n+1}$, for the
exchanged partons was naturally included into the evolution equations.
Moreover, the evolution \dl was included into the equation kernels, and the
nonperturbative input was separated as an inhomogeneous term. This allowed
for including the kernels  of the standard DGLAP evolution into the equations,
which was necessary, in order to obtain a proper behaviour of the 
parton distributions at moderate and large values of $x$.

The full (unified) equation for the nonsinglet unintegrated parton distribution,
$f_{NS}$ was:
\eqn
f_{NS}(x,Q^2)&=&\as(\dt P \ot \dt q_{NS}^{(0)})(x)
+\as\int_{k_0^2}^{Q^2} {dk^2 \over k^2}\,(\dt P \ot f_{NS})(x,k^2)\nonumber\\
&&\hspace*{10ex}{\bf(\hspace*{3ex}DGLAP\hspace*{3ex})}\nonumber\\
&+&
\as{4\over 3}
\int_{x}^1 {dz\over z}
\int_{Q^2}^{Q^2/z}
{dk^{2}\over k^{2}}
f_{NS}\left({x\over z},k^{2}\right)\nonumber\\
&&\hspace*{10ex}{\bf (\hspace*{3ex}Ladder\hspace*{3ex})}\nonumber\\
&-&\as
\int_{x}^1 {dz\over z}
\Biggl(
\Biggl[ \frac{\tilde  {\bf F}_8 }{\omega^2} \Biggr](z)
\frac{ {\bf G}_0 }{2\pi^2}
\Biggr)_{qq}
\int_{k_0^2}^{Q^2}
{dk^{2}\over k^{2}}
f_{NS}\left({x\over z},k^{2}\right)\nonumber\\
&-&\as
\int_{x}^1 {dz\over z}
\int_{Q^2}^{Q^2/z}
{dk^{2}\over k^{2}}
\Biggl(
\Biggl[\frac{\tilde   {\bf F}_8 }{\omega^2} \Biggr]
\Biggl(\frac{k^{2}}{Q^2}z \Biggr)\frac{ {\bf G}_0 }{2\pi^2}
\Biggr)_{qq}
f_{NS}\left({x\over z},k^{2}\right).\nonumber\\
&&\hspace*{10ex}{\bf(Nonladder)}
\label{nloinf}
\eqnx
For a detailed form of the kernels see
Appendix A. Matrices, ${\bf F}_8$ and ${\bf G}_0$, represent 
the octet partial waves and the colour factors respectively. 
They are described in a detail in Appendix A.
Symbol $\displaystyle \Biggl[{\widetilde { {\bf F}_8}/{\omega^2} }\Biggr](z)$
denotes the inverse Mellin transform of
$\displaystyle {{\bf F}_8}/{\omega^2}$~:
\eq
\Biggl[{\widetilde { {\bf F}_8}/{\omega^2}  }\Biggr](z)=
\int_{\delta-i\infty}^{\delta+i\infty} {d\omega \over 2\pi i}
z^{-\omega}{{\bf F}_8(\omega)}/{\omega^2}.
\label{imellin}
\eqx	
with the integration contour located to the right of the singularities
of the function $\displaystyle {\bf F}_8(\omega)/{\omega^2}$.

The singlet distribution, $f_S=\ve{f_{\Sigma}}{f_g}$,
was resummed by a similar vector equation,
formulated for the quark and the gluon components \cite{kz}.

The integrated parton distributions, $\dt q_i(x,Q^2)$, were
obtained with the unintegrated ones, using the relation:
\eq
\dt q_i(x,Q^2)= \dt q_i^{(0)}(x)+\int_{k_0^2}^{W^2}\,{dk^2 \over k^2}\,
f_i(x^{\prime}=x(1+{k^2\over Q^2}),k^2),
\label{fix}
\eqx
where the phase-space was extended from $Q^2$ to $W^2=Q^2(1/x-1)$
corresponding to the total energy squared measured in the center-of-mass
frame \cite{BBJK}. The index, $i=S,NS,g$, and $\dt q_i^{(0)}(x)$ was the
input parton distribution including the contributions from the
nonperturbative region.

Finally, the structure function, $g_1$, at the LO accuracy was:
\eqn
g_1(x,Q^2)={\cor{e^2} \over 2} \left\{ \dt q_{NS}(x,Q^2)
+\dt q_S(x,Q^2) \right\},
\label{g1jet}
\eqnx
where $\langle e^2\rangle = {1\over N_f}\sum_{l=1}^{N_f}e_l^2$, 
and $N_f$ denoted the number of active flavours ($N_f=3$).

\begin{figure}[htb]
   \vspace*{-1cm}
    \centerline{
     \psfig{figure=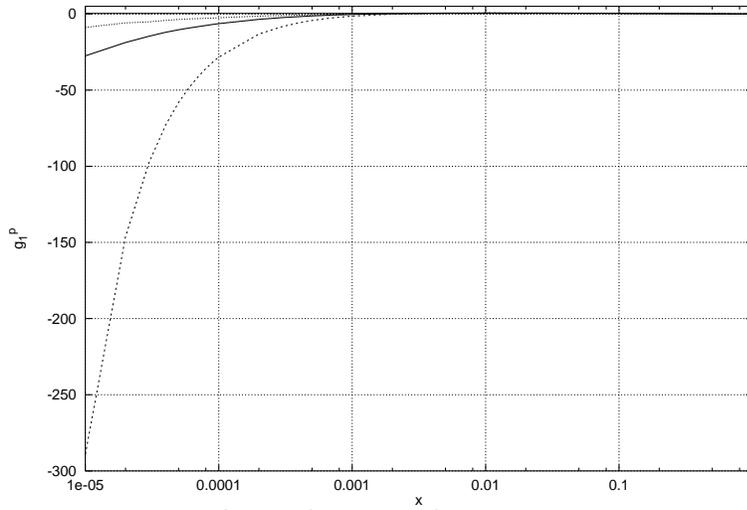,width=10cm}
               }
    \vspace*{-0.5cm}
\caption{\footnotesize{ Structure function, $g_1^p(x,Q^2)$, at $Q^2=10$ GeV$^2$ plotted as
a function of $x$.
Solid line corresponds to the results with the full
$ln^2(1/x)$ resummation, where the nonladder terms, ladder terms and the
DGLAP kernels were included. Dotted line shows the pure DGLAP evolution.
Thin solid line shows the nonperturbative input, $g_1^{(p, 0)}$, and
dashed line shows the incomplete DL resummation, where the nonladder
terms were neglected and the DGLAP terms were included.}}
\label{fig1}
\end{figure}

The analysis of $g_1$ obtained with the unified evolution including the DGLAP
terms at the LO accuracy was performed in \cite{kz}. The results obtained
with
the standard LO DGLAP evolution and the unified evolution (DL+LO DGLAP) were similar
at $x>10^{-3}$. Significant discrepancies appeared at very low $x$, $x<10^{-4}$,
when the corrections, \dl, were predominant (Fig.\ \ref{fig1}).

\begin{figure}[htb]
   \vspace*{-1cm}
    \centerline{
     \psfig{figure=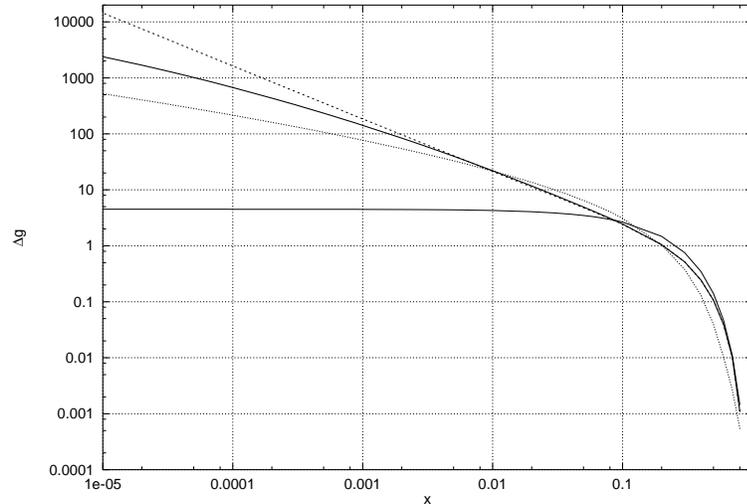,width=10cm}
               }
    \vspace*{-0.5cm}
     \caption{\footnotesize{ The spin dependent gluon distribution, $\Delta g(x,Q^2)$,
at $Q^2=10$ GeV$^2$ plotted as a function of $x$.
Solid line corresponds to the results with the full
$ln^2(1/x)$ resummation, where the nonladder terms, ladder terms and the
DGLAP kernels were included. Dotted line shows the pure DGLAP evolution.
Thin solid line shows the nonperturbative input, $\Delta g^{(0)}$, and
dashed line shows the incomplete DL resummation, where the nonladder
terms were neglected and the DGLAP terms were included.}}
\label{fig2}
\end{figure}

For the gluon distribution, $\Delta g$, the discrepancies between the LO DGLAP 
and DL+LO DGLAP curves were large (Fig.\ \ref{fig2}).
This shows that the unified evolution of gluon distribution is driven
by the ladder and the nonladder terms.

Recently, the unified equation were completed by the NLO DGLAP terms
\cite{ziaja1,ziaja2}.
The relations between the NLO DGLAP and the DL+NLO DGLAP curves of
Refs.\ \cite{ziaja1,ziaja2} were similar as the relations between the
LO DGLAP and the DL+NLO DGLAP curves obtained in \cite{kz}.


\section{Spin structure of the polarized photon}


Unified evolution equations may be also applied to investigate the structure
of the polarized photon in a deep inelastic scattering of an electron on photon
beams \cite{foton}. The spin structure function of the polarized photon, $g_1^{\gamma}$,
will be measured with future linear colliders $e^+e^-$ and $e\gamma$
\cite{STRATMANN1,phot1,phot2}. In particular, the scattering $e\gamma$ will
probe the $g_1^{\gamma}$
at low values of Bjorken $x$. The partonic content of the polarized photon
can be also measured with the process of the dijet photoproduction
at the scattering of an electron on a proton \cite{phot3}.

Using the unified equations for the description of $e\gamma$ scattering
requires an extra inhomogeneous term, $\dt k_i(x,Q^2)$, to be included into
these equations. This term describes the pointlike coupling  of the photon
to quarks, antiquarks and gluons \cite{STRATMANN1}. It joins the DGLAP evolution
equations in the following way:
\eqn
\ddq \dt q_{i}(x,Q^2)&=&\dt k_i(x,Q^2) + \as\,(\dt P \ot \dt q_{i})(x,Q^2),
\label{diff}
\eqnx
where $i=S,NS,g$. There is no photon-gluon coupling at the LO of DGLAP,
$\dt k_g(x,Q^2)=0$ \cite{STRATMANN1}.

In Ref.\ \cite{foton} we found an analytic solution of the simplified evolution
equations. They described the DL evolution in an approximation, where
the nonladder terms were neglected.
We got the asymptotic form of $g_1^{\gamma}$ at the limit, $x\rightarrow0$.
As expected, $g_1^{\gamma}$ scaled with $x$. The scaling coefficient was
negative, and it was related in a simple way to the asymptotic scaling coefficients
of the nucleon structure function, $g_1$.

Then we found a numerical solution for $g_1^\gamma$ with the full evolution 
equations,
taking into account the standard DGLAP evolution at the LO accuracy. The
input parametrization fulfilled the sum rule for $g_1^\gamma$
\cite{bass1,bass2}:
\eq
\int_0^1\,dx\,g_1^{\gamma}(x,Q^2)=0.
\label{g1gamma}
\eqx
We considered two limiting cases for the input parametrization of the quark
and gluon distributions:(i) the case, when $\dt q^{(0)}=0$ and
$\dt g^{(0)}=0$, and the solutions of the evolution equations were generated
radiatively from the inhomogeneous terms, $\dt k$, (ii) the case, when
$\dt q^{(0)}=0$ but the gluon input was nonzero. In that case the input
paraterization of gluon distribution was obtained with the
vector-meson-dominance model (VMD), assuming the dominance of the mesons,
$\rho$ and $\omega$ \cite{vmd1,vmd2}.

\begin{figure}[t]
\centerline{\epsfig{figure=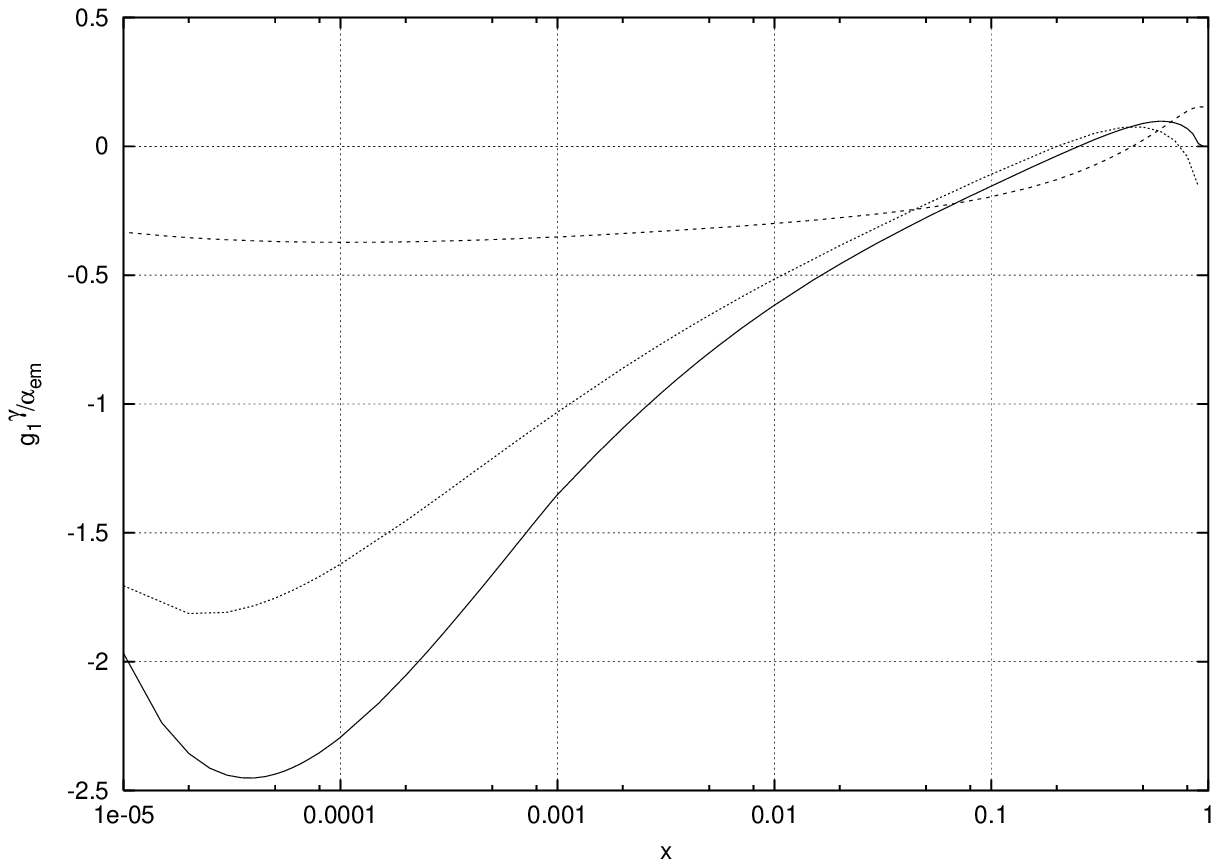,width=10cm}}
\caption{{\footnotesize Structure function, $\gp(x,Q^2)/\alpha_{em}$,
at $Q^2=10$ GeV$^2$ obtained after solving the unified evolution equations
with the input parametrization (i). Solid line corresponds to the results with
the full $ln^2(1/x)$ resummation, where the nonladder, ladder corrections
and the LO DGLAP kernels were included. Dashed line shows the LO DGLAP evolution,
and dotted line shows the NLO DGLAP evolution.}}
\label{fig3}
\end{figure}

\begin{figure}[t]
\centerline{\epsfig{figure=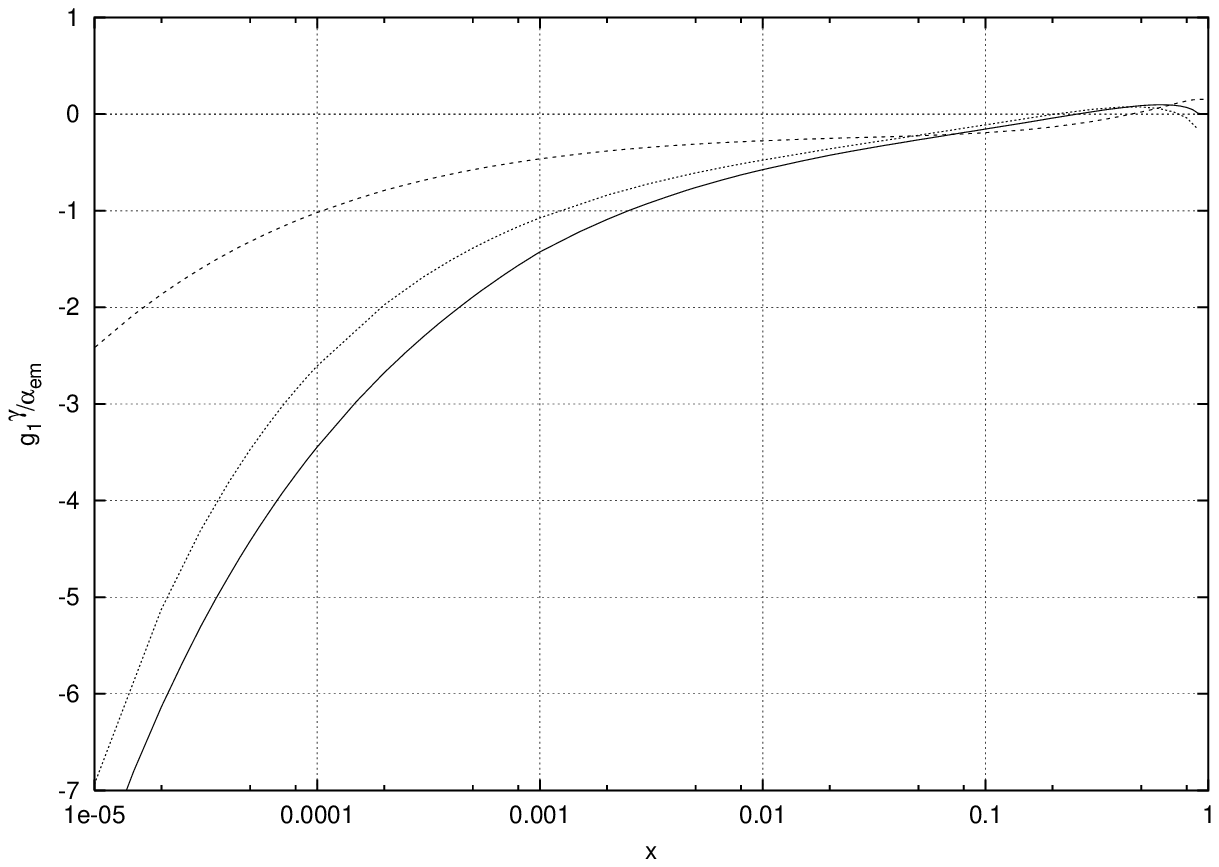,width=10cm}}
\caption{{\footnotesize Structure function, $\gp(x,Q^2)/\alpha_{em}$,
at $Q^2=10$ GeV$^2$ obtained after solving the unified evolution equations
with the input parametrization (ii). Solid line corresponds to the results with
the full $ln^2(1/x)$ resummation, where the nonladder, ladder corrections
and the LO DGLAP kernels were included. Dashed line shows the LO DGLAP evolution,
and dotted line shows the NLO DGLAP evolution.}}
\label{fig4}
\end{figure}

Including the double logarithmic corrections significantly influcenced
the behaviour of $g_1^{\gamma}$. The differences between the
curves obtained with the LO DGLAP, NLO DGLAP and DL+LO DGLAP evolution strongly
depended on the input parametrization of the parton distributions used (Figs.\
\ref{fig3},\ref{fig4}).
It depended also on the infrared cut-off, $k_0^2$ (not shown). 
However, the estimated
value of asymmetry, $g_1^{\gamma}/F_1^{\gamma}$ was small, $\sim 10^{-3}$ at
$x=10^{-4}$, i.e. it was difficult to measure.

\begin{figure}[t]
\centerline{\epsfig{figure=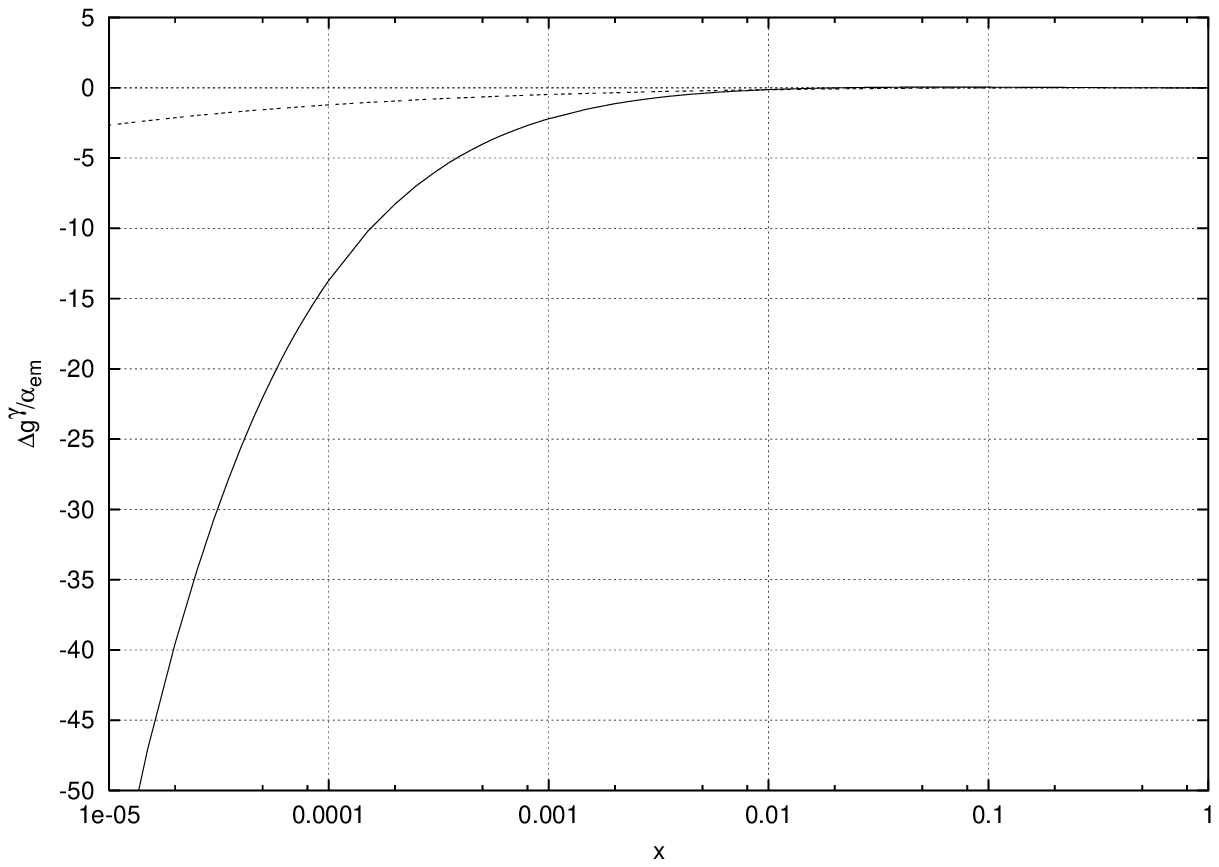,width=10cm}}
\caption{{\footnotesize Spin dependent gluon distribution in photon,
$\Delta g^{\gamma}(x,Q^2)/\alpha_{em}$,
at $Q^2=10$ GeV$^2$ obtained after solving the unified evolution equations
with the input parametrization (i).
Solid line corresponds to the results obtained with
the full $ln^2(1/x)$ resummation, where the nonladder, ladder corrections
and the LO DGLAP kernels were included, dashed line shows the LO DGLAP 
evolution.}}
\label{fig5}
\end{figure}

\begin{figure}[t]
\centerline{\epsfig{figure=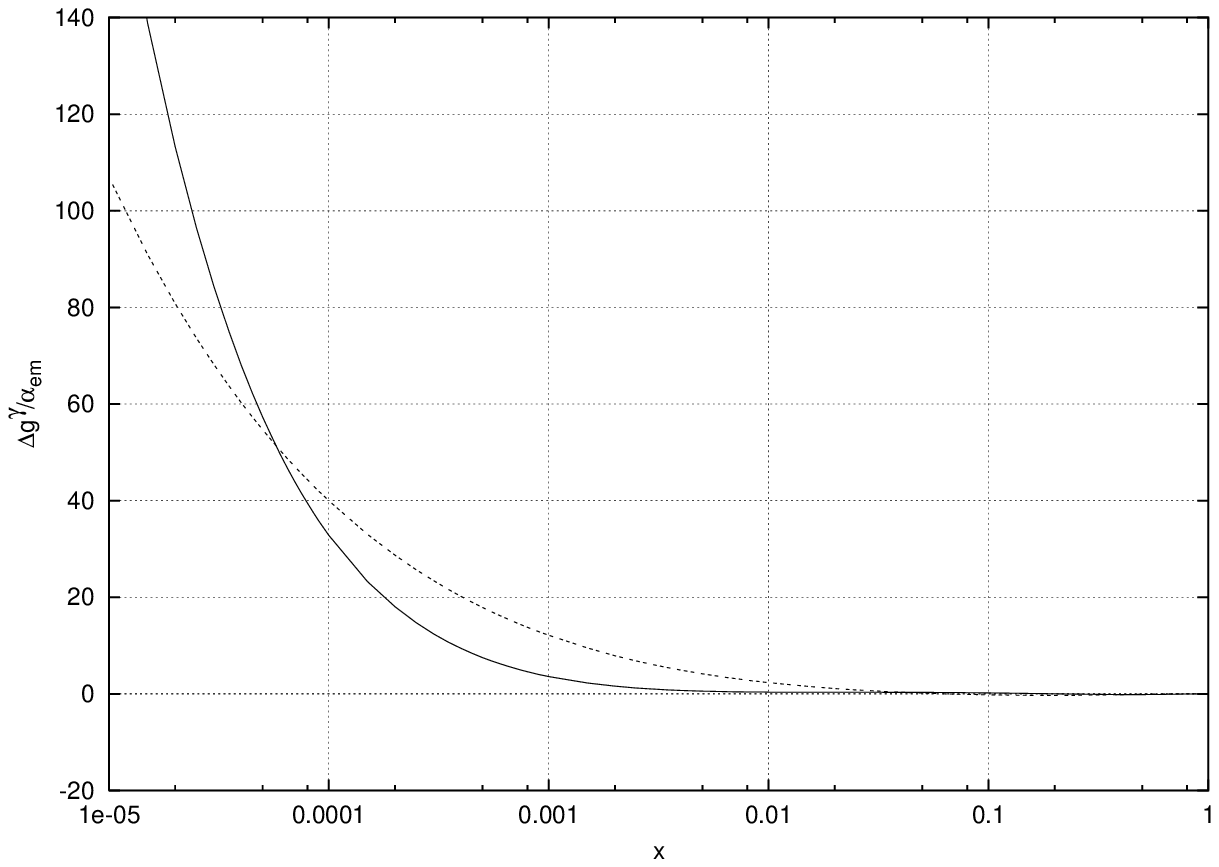,width=10cm}}
\caption{{\footnotesize Spin dependent gluon distribution in photon,
$\Delta g^{\gamma}(x,Q^2)/\alpha_{em}$,
at $Q^2=10$ GeV$^2$ obtained after solving the unified evolution equations
with the input parametrization (ii).
Solid line corresponds to the results obtained with
the full $ln^2(1/x)$ resummation, where the nonladder, ladder corrections
and the LO DGLAP kernels were included, dashed line shows the LO DGLAP 
evolution.}}
\label{fig6}
\end{figure}

The gluon distribution, $\dt g$, obtained strongly depended on the input
parametrization used (Figs.\ \ref{fig5},\ref{fig6}). In case (i) 
$\dt g$ was negative at low values of $x$, in case (ii) it was positive. 
Future experiments will be then able to determine which parametrization 
is correct.


\section{Logarithmic corrections tested}

The presence and the magnitude of the logarithmic corrections, \dl, can be
tested in the polarized deep inelastic scattering accompanied by a forward
jet. This idea was first applied for testing the BFKL resummation in the
unpolarized DIS \cite{mueller1,mueller2}.

Assume, that a forward jet is produced in the DIS of an electron on a
proton. The momentum coordinates of the jet are, $(x_J,k_J^2)$, where
$x_J$ is the longitudinal momentum fraction, and $k_J^2$ denotes the
transverse momentum of the jet squared. The momentum transfer carried by
the photon is $q$, and the Bjorken variable $x$ is defined as:
$x=Q^2/(2pq)$, where $Q^2=-q^2$ and $p$ is the momentum of the proton.

If the jet coordinates fulfill the requirement:
\eqn
x_J&\gg&x,\nonumber\\
k_J^2&\sim&Q^2,
\label{jet}
\eqnx
the jet is produced at low values of $x/x_J$. A contribution to this
process coming from the standard DGLAP evolution is then supressed
($k_J^2\sim Q^2$) \cite{mueller1,mueller2,supp1,supp2,supp3,supp4}, 
and the nonperturbative region is 
unpenetrated by the evolution, $\ln^2(1/x)$ \cite{supp5,supp6}.

In this process we measure the differential cross section. It is related to
the momentum transfer, $Q^2$, the electron energy fraction, $y$, and the
differential structure function $x_J\partial^2g_1/\partial x_J \partial
k_J^2$. In analogy to the full structure function, $g_1$, the differential
structure function may be expressed through the integrated parton
distributions in a proton and the unintegrated distributions of quarks and
antiquarks in a parton \cite{fjet1}. 

\begin{figure}[t]
    \centerline{
     \epsfig{figure=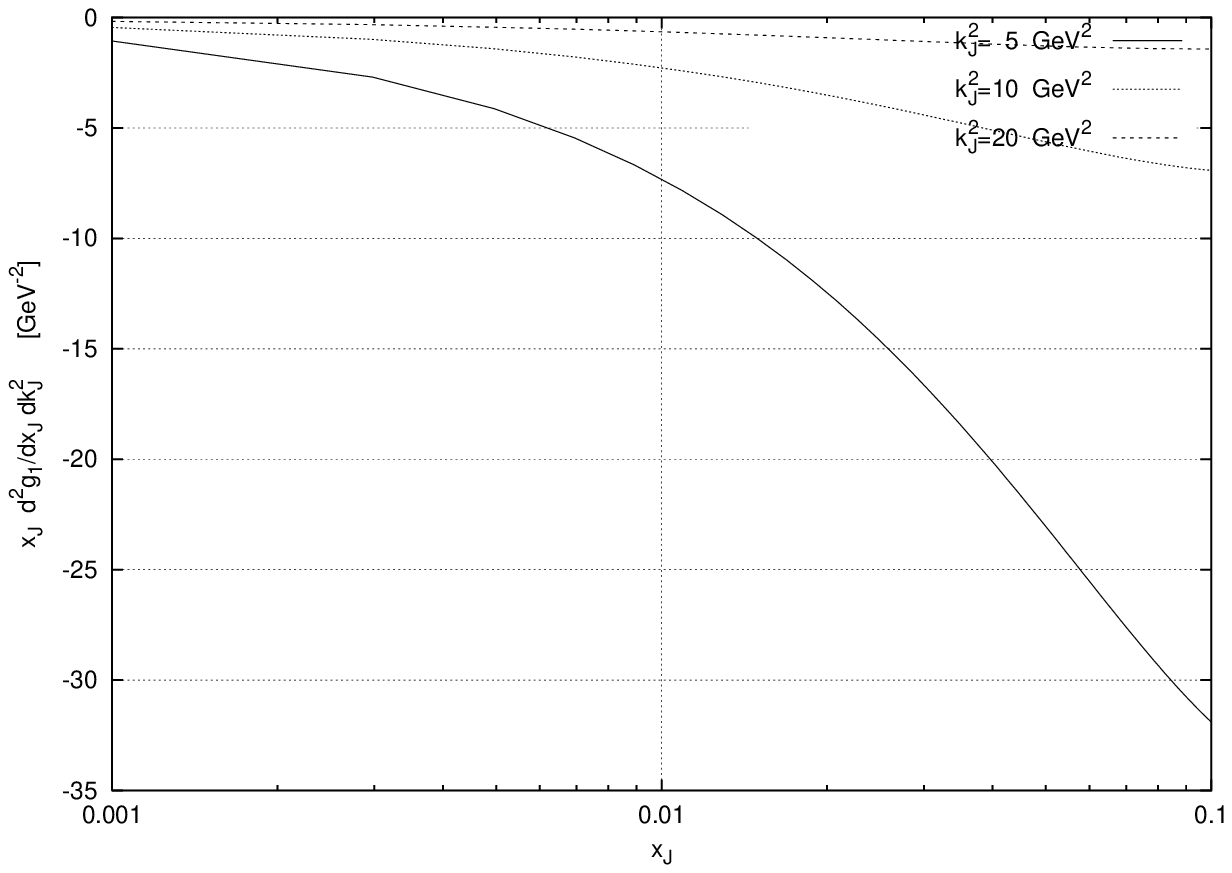,width=10cm}
               }
\caption{{\footnotesize The differential spin structure function,
$x_J{\partial g_1\over \partial x_J \partial k_J^2}$, for the fixed
$\bar\alpha_s$ (case (i)) plotted as a function of the
longitudinal momentum fraction carried by a jet, $x_J$.
We show predictions for the three different values of the transverse momentum
of the jet squared, $k_J^2=$ 5 GeV$^2$, 10 GeV$^2$ and 20 GeV$^2$.
Those calculations were performed at $Q^2=10$ GeV$^2$ and $x=10^{-4}$.}}
\label{fig7}
\end{figure}

\begin{figure}[t]
    \centerline{
     \epsfig{figure=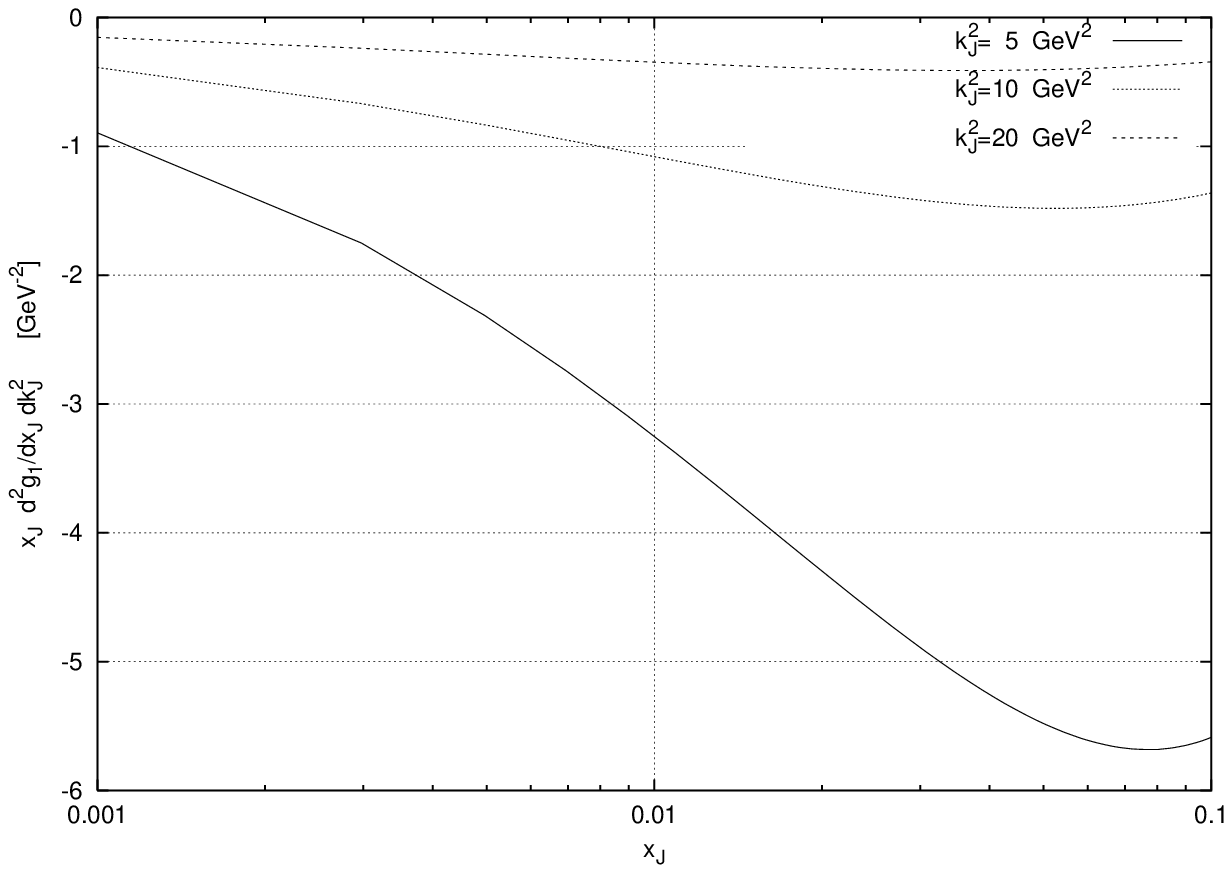,width=10cm}
               }
\caption{\footnotesize{The differential spin structure function,
$x_J{\partial g_1\over \partial x_J \partial k_J^2}$, for the running
$\bar\alpha_s$ (case (ii)) plotted as the function of the longitudinal
momentum fraction carried by a jet, $x_J$.
We show predictions for the three different values of the transverse
momentum of the jet squared, $k_J^2=$5 GeV$^2$,
10 GeV$^2$ and 20 GeV$^2$.  Those  calculations
were performed  at $Q^2=10$ GeV$^2$ and $x=10^{-4}$.}}
\label{fig8}
\end{figure}

In Ref.\ \cite{fjet1} we formulated the \dl evolution equations for the unintegrated
distributions of (anti)quarks in a parton. In those equations we included
the leading ladder terms, and neglected the subleading ones. We neglected
also the DGLAP terms as the DGLAP evolution was supressed \cite{fjet1}.
We found an analytic solution of those equations in the following cases: (i)
for the fixed coupling constant, $\alpha_s(\mu^2)$, where
$\mu=(k_J^2+Q^2)/2$ \cite{supp1,supp2}, (ii) for the running coupling constant,
$\alpha_S(\mu^2)$, where $\mu^2=k_f^2/\zeta$, and $k_f^2$ denotes the
transverse momentum of (anti)quark squared, and $\zeta$ is its longitudinal
momentum fraction. After numerical integrating the unintegrated
distributions we got the differential structure function,
$x_J\partial^2g_1/\partial x_J \partial k_J^2$ (Figs.\ \ref{fig7},\ref{fig8}). 
This function strongly
changed with the value of the ratio, $x/x_J$, and it depended also on the
transverse momentum of the jet, $k_J^2$. The strong dependence on $x/x_J$
was a direct effect of the logarithmic resummation. Comparing the results, we
found that the case (ii) with the running coupling constant was
more realistic than the case (i) with the fixed coupling constant. In both cases
the \dl effects were much larger than the background described by the Born
approximation (not shown).

\begin{figure}[t]
\centerline{\epsfig{figure=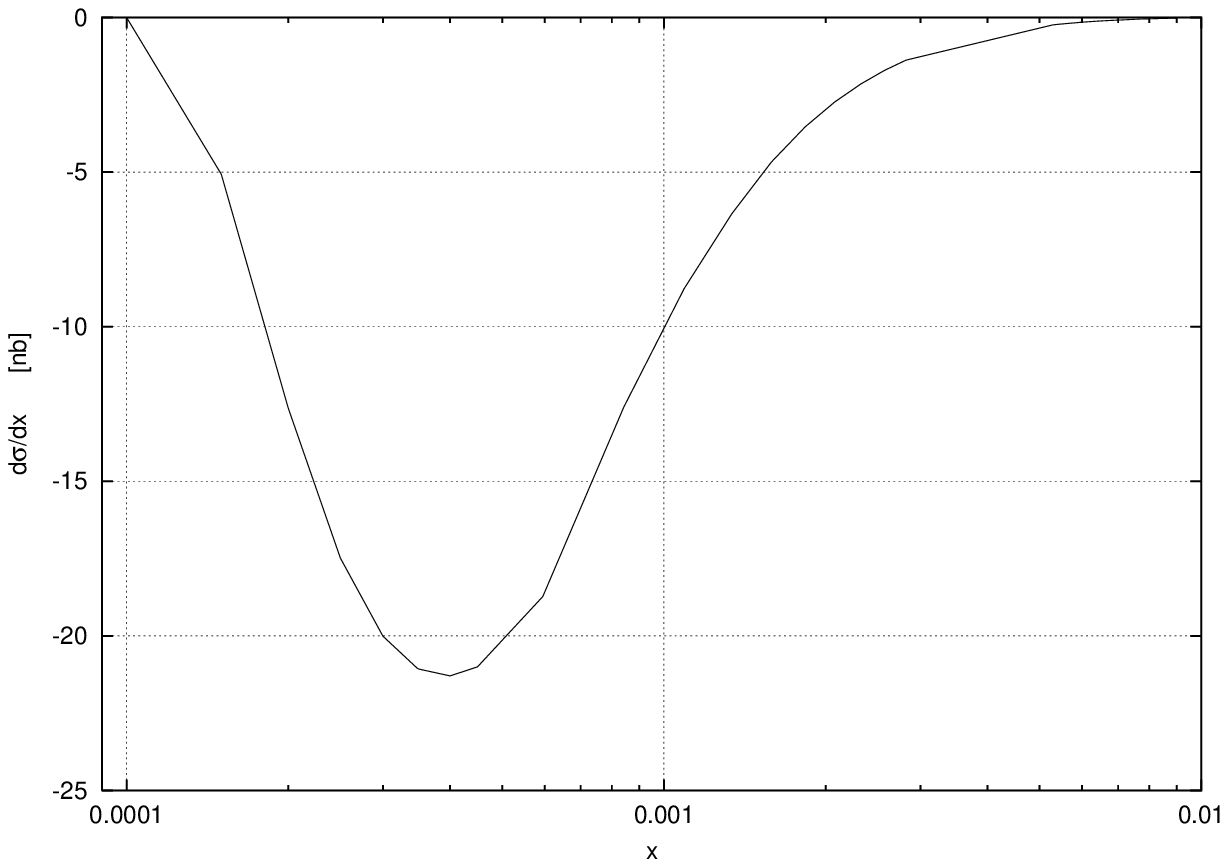,height=6cm,width=9cm}}
\centerline{\epsfig{figure=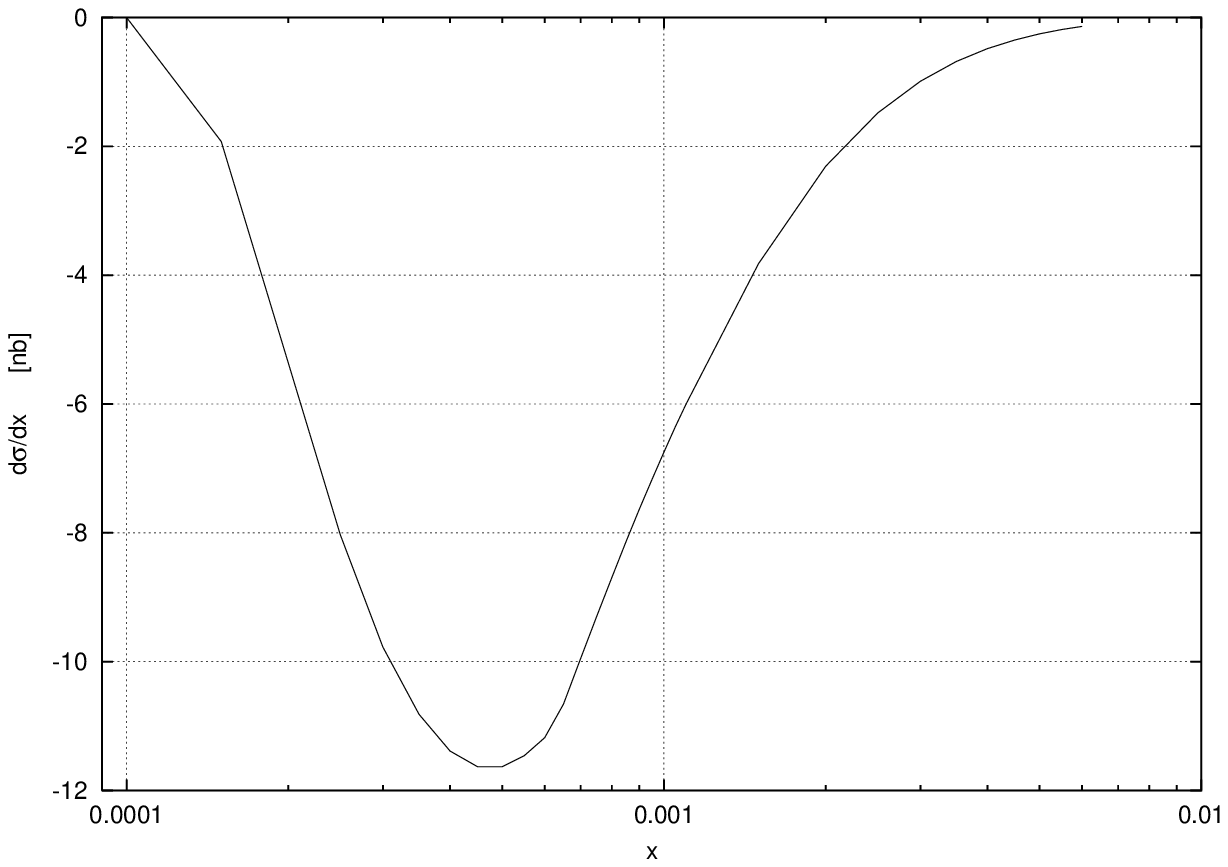,height=6cm,width=9cm}}
\centerline{\epsfig{figure=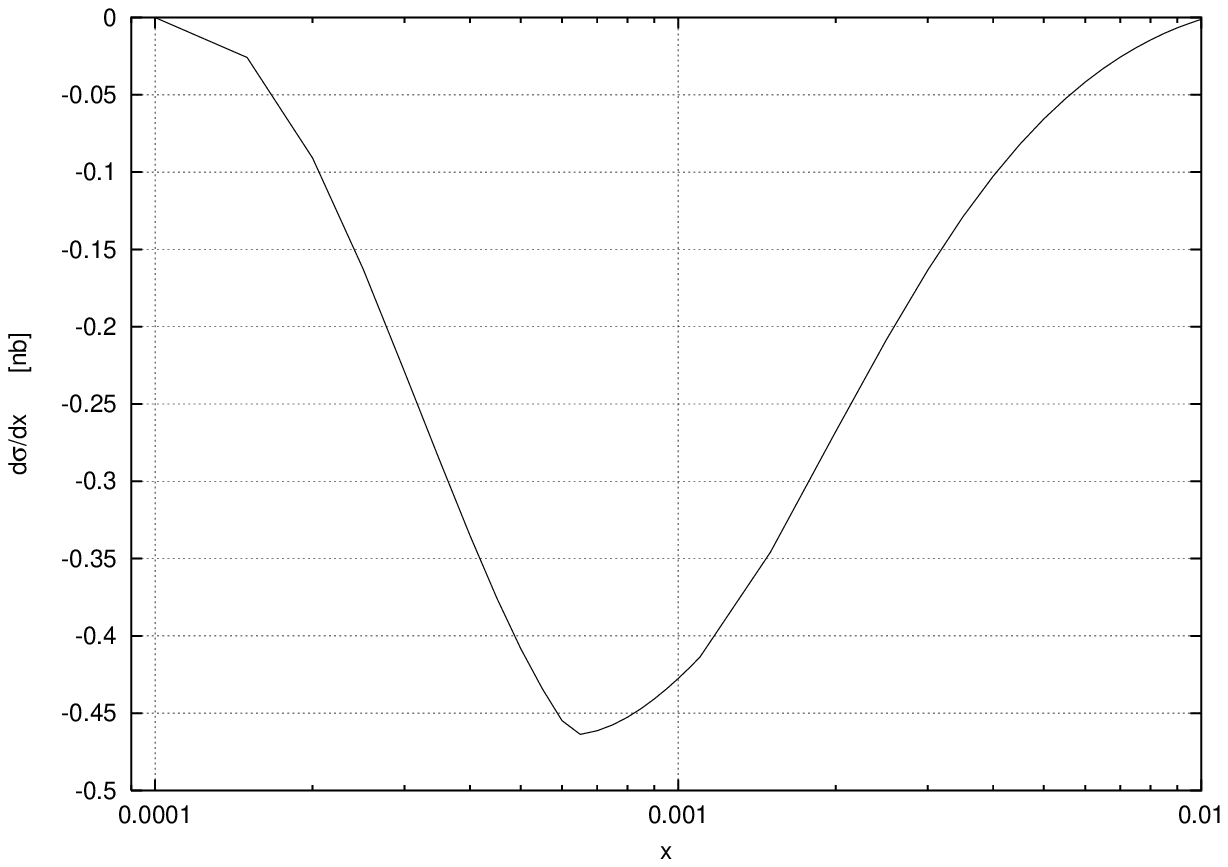,height=6cm,width=9cm}}
\caption{\footnotesize{
The cross-section, ${d\sigma \over dx}$, for the forward jet production
in the polarized deep inelastic scattering.
Figures 2a and 2b show the results with the double
logarithmic $ln^2(1/\xi)$ effects included. They correspond to two
choices of the scale $\mu^2$: (i) $\mu^2=(k_J^2 + Q^2)/2 $ (Fig.\ 2a),
(ii) $\mu^2=k_f^2/\xi$ (Fig.\ 2b).
Fig.\ 2c shows the cross-section  ${d\sigma \over dx}$ in the Born
approximation where the double logarithmic resummation was neglected.}}
\label{fig9}
\end{figure}

In Ref.\ \cite{fjet2} we estimated the cross section and the asymmetry for the DIS
accompanied by a forward jet. We used the kinematical cuts as planned for
the future collider, the polarized HERA \cite{dane1,dane2,dane3}. The effects of \dl resummation
contributed significantly to the asymmetric cross section $d\sigma/dx$,
where $x$ was the Bjorken variable (Fig.\ \ref{fig9}). The asymmetry obtained 
was, however, small, and changed between $-0.01$ and $-0.04$ at low $x$.


\section{Summary}

We performed the analysis of the polarized deep inelastic scattering in the
region of low values of Bjorken $x$. We used the evolution equations
formulated for the unintegrated parton distributions. Those equations
included the complete resummation of the double logarithmic contributions,
$\ln^2(1/x)$, and the standard LO DGLAP evolution terms. The DGLAP evolution
was included into those equations, in order to obtain the proper behaviour 
of the parton distributions at the moderate and large values of $x$.

Using that formalism, we obtained predictions for the following observables:
(i) the nucleon structure function, $g_1$, in the polarized
deep inelastic scattering, (ii) the structure function of the polarized
photon, $g_1^{\gamma}$, in the scattering of a lepton on a polarized photon,
and (iii) the differential structure function, 
$x_J\,\partial^2g_1/\partial x_J\,\partial k_J^2$,
in the polarized deep inelastic scattering accompanied by a forward jet.

For each process the consequences of including the logarithmic corrections
were studied in a detail. Generally,
we observed an enhancement of the magnitude of $g_1^{p}$, $g_1^{\gamma}$  
and $\dt g$
obtained with the unified evolution as compared to the pure LO DGLAP results.
Especially, the behaviour of gluon distributions with the unified evolution
at low $x$ was clearly dominated by the large contribution of the ladder and
nonladder terms. Some predictions for the observables, the asymmetry and
the cross sections, in the processes (i)-(iii) were also obtained.
The case (iii) tested the presence and the magnitude of the $\ln^2(1/x)$
contributions in DIS. In that case the asymmetry parameter estimated with the
kinematical cuts of the future collider, polarized HERA, was between
$-0.04$ and $-0.01$ at low $x$.

To sum up, our observations suggest that the standard DGLAP evolution is
not complete at low $x$, where the effects of $ln^2(x)$ resummation are large
and therefore cannot be neglected.
This result is important to planned experiments with the polarized HERA and
linear colliders, which would probe the region of low values of Bjorken
$x$.

\section*{Acknowledgements}

I would like to express my deep gratitude to Professor Jan Kwieci\'nski
who introduced me into the spin physics, who inspired and supported my research 
in this field. I learned from him so much.\\

\nin
This research has been supported in part by the Polish State Committee for
Scientific Research with grants 2 P03B 05119, 2PO3B 14420,  and European
Community grant 'Training and Mobility of Researchers', Network 'Quantum
Chromodynamics and the Deep Structure of Elementary Particles'
FMRX-CT98-0194.

\section*{Appendix A}

Here a brief description of the evolution kernels of Eq.\ \re{nloinf}
is given. DGLAP kernels were taken from Ref.\ \cite{reya}. Here the DGLAP kernel
$\dt P$ includes only the LO terms:
\eq
\dt P=\dt P^{(0)}.
\label{dglap}
\eqx

The ladder kernels corresponding to the LO DGLAP kernels at the longitudinal
momentum transfer, $z=0$, \cite{kz}
generate the double logarithmic corrections in the region of $Q^2<k^2<Q^2/z$.

The nonladder kernels were obtained in Ref.\ \cite{kz} from the infrared evolution
equations written for the singlet partial waves ${\bf F_0}$, $\bf F_8$
\cite{BARTNS,BARTS,QCD,QCD1}. In \cite{kz} we noticed that extending
the kernel of the double logarithmic evolution equations from the ladder one,
\eq
\as \dt P_{qq}/\omega\nonumber,
\eqx
to the modified one,
\eq
\as \left( \dt P_{qq}/\omega
-({\bf F_8}(\omega)\,{\bf G_0})_{qq}/(2\pi^2\omega^2) \right)\nonumber,
\eqx
gave a proper anomalous dimension as derived from the infrared
evolution equations.

Matrix ${\bf G}_0$ contained colour factors resulting from attaching the soft
gluon to external legs of the scattering amplitude~:
\eqn
{\bf G}_0 &=&\left( \begin{array}{cc}  {N^2-1 \over 2N} & 0  \\
                                                    0 & N  \\ \end{array}
						    \right ),
\label{g0}
\eqnx
where $N$ was the number of colours.

Further, it was checked that the Born approximation of ${\bf F_8}$,
\eq
{\bf F}_8^{Born}(\omega)\approx 8\pi^2 \as \frac{{\bf M}_8}{\omega}.
\eqx
gave accurate results for the DL evolution.
Martix ${\bf M}_8$ was a splitting function matrix in the
colour octet $t-$channel,
\eqn
{\bf M}_8 &=&\left( \begin{array}{cc} -{1 \over 2N} & -{N_F \over 2}\\
                                                N & 2N \\ \end{array} \right ).
\eqnx
The inverse Melin transform of ${\bf F}_8^{Born}(\omega)$ then read~:
\eq
\Biggl[\frac{\tilde {\bf F}_8^{Born}}{\omega^2}\Biggr](z)=
4\pi^2 \as {\bf M}_8 ln^2 (z).
\label{born}
\eqx

The evolution equation \re{nloinf} includes the nonladder corrections in
the Born approximation \re{born}.

\begin{thebibliography}{10}

\bibitem{reya}
B.~Lampe and E.~Reya.
\newblock {\em Phys.Rept.}, 332:1, 2000.

\bibitem{jet3}
H.-Y. Cheng.
\newblock {\em Chin. J. Phys}, 38:753, 2000.

\bibitem{vetterli}
M.~C. Vetterli.
\newblock {\em {hep-ph/9812420}}.

\bibitem{ioffe}
B.~L. Ioffe.
\newblock {\em Surveys High Energ. Phys.}, 8:107, 1995.

\bibitem{pemc}
J.~Ashman {et al.}
\newblock {EMC}.
\newblock {\em Phys. Lett. B}, 206:364, 1988.

\bibitem{vann1}
R.~Mertig and W.~L. van Neerven.
\newblock {\em Z. Phys. C}, 70:637, 1996.

\bibitem{vogel}
W.~Vogelsang.
\newblock {\em Phys. Rev. D}, 54:2023, 1996.

\bibitem{axial}
R.~D. Carlitz, J.~C. Collins, and A.~H. Mueller.
\newblock {\em Phys.\ Lett.\ B}, 214:229, 1988.

\bibitem{leader2}
E.~Leader, A.~V. Sidorov, and D.~B. Stamenov.
\newblock {\em Phys. Lett. B}, 445:232, 1998.

\bibitem{BARTNS}
J.~Bartels, B.~I. Ermolaev, and M.~G. Ryskin.
\newblock {\em Z. Phys. C}, 70:273, 1996.

\bibitem{BARTS}
J.~Bartels, B.~I. Ermolaev, and M.~G. Ryskin.
\newblock {\em Z. Phys. C}, 72:627, 1996.

\bibitem{blum9603}
J.~Bl\"umlein and A.~Vogt.
\newblock {\em Acta Phys. Pol. B}, 27:1309, 1996.

\bibitem{blum9606}
J.~Bl\"umlein and A.~Vogt.
\newblock {\em Phys. Lett. B}, 386:350, 1996.

\bibitem{kiyo}
Y.~Kiyo, J.~Kodaira, and H.~Tochimura.
\newblock {\em Z.\ Phys.\ C}, 74:631, 1997.

\bibitem{BBJK}
B.~Bade\l{}ek and J.~Kwieci\'nski.
\newblock {\em Phys. Lett. B}, 418:229, 1998.

\bibitem{kz}
J.~Kwieci\'nski and B.~Ziaja.
\newblock {\em Phys. Rev. D}, 60:054004, 1999.

\bibitem{foton}
J.~Kwieci\'nski and B.~Ziaja.
\newblock {\em Phys. Rev. D}, 63:054022, 2001.

\bibitem{fjet1}
J.~Kwieci\'nski and B.~Ziaja.
\newblock {\em Phys. Lett. B}, 464:293, 1999.

\bibitem{fjet2}
J.~Kwieci\'nski and B.~Ziaja.
\newblock {\em Phys. Lett. B}, 470:247, 1999.

\bibitem{regg1}
B.~L. Ioffe, V.~A. Khoze, and L.~N. Lipatov.
\newblock {Hard Processes}.
\newblock {\em {North-Holland, Amsterdam}}, 1984.

\bibitem{regg2}
J.~Ellis and M.~Karliner.
\newblock {\em Phys. Lett. B}, 213:73, 1988.

\bibitem{bfkl1}
E.~A. Kuraev, L.~N. Lipatov, and V.~Fadin.
\newblock {\em Sov. Phys. JETP}, 45:199, 1977.

\bibitem{bfkl2}
Y.~Y. Balitsky and L.~N. Lipatov.
\newblock {\em Sov. J. Nucl. Phys.}, 28:822, 1978.

\bibitem{QCD}
R.~Kirschner and L.~N. Lipatov.
\newblock {\em Nucl. Phys. B}, 213:122, 1983.

\bibitem{QCD1}
R.~Kirschner.
\newblock {\em Z. Phys. C}, 67:459, 1995.

\bibitem{ziaja1}
B.~Ziaja.
\newblock {\em Phys. Rev. D}, 66:114017, 2002.

\bibitem{ziaja2}
B.~Ziaja.
\newblock {\em to be published in Eur. Phys. J. C}, 2003.

\bibitem{STRATMANN1}
M.~Stratmann and W.~Vogelsang.
\newblock {\em Phys. Lett. B}, 386:370, 1996.

\bibitem{phot1}
M.~Stratmann.
\newblock {\em Nucl. Phys. B (Proc. Suppl.)}, 82:400, 2000.

\bibitem{phot2}
A.~Vogt.
\newblock {\em Nucl. Phys. B (Proc. Suppl.)}, 82:394, 2000.

\bibitem{phot3}
M.~Stratmann and W.~Vogelsang.
\newblock {Towards the parton densities of polarized photons at HERA}.
\newblock {\em {*Hamburg 1999, Polarized protons at high energies - Accelerator
  challenges and physics opportunities*}}, page 324, 1999.

\bibitem{bass1}
S.~D. Bass.
\newblock {\em Int. J. Mod. Phys. A}, 7:6039, 1992.

\bibitem{bass2}
S.~Narison, G.~M. Shore, and G.~Veneziano.
\newblock {\em Nucl. Phys. B}, 391:69, 1993.

\bibitem{vmd1}
J.~J. Sakurai.
\newblock {\em Ann. Phys.}, 11:1, 1960.

\bibitem{vmd2}
T.~T.~Bauer et~al.
\newblock {\em Rev. Mod. Phys.}, 50:261, 1978.

\bibitem{mueller1}
A.~Mueller.
\newblock {\em Nucl. Phys. (Proc. Suppl.)}, 18C:125, 1990.

\bibitem{mueller2}
A.~Mueller.
\newblock {\em J. Phys. G}, 17:1443, 1991.

\bibitem{supp1}
J.~Bartels, A.~De Roeck, and M.~Loewe.
\newblock {\em Z. Phys. C}, 54:635, 1992.

\bibitem{supp2}
W.~K. Tang.
\newblock {\em Phys. Lett. B}, 278:363, 1992.

\bibitem{supp3}
J.~Kwieci\'nski, A.~D. Martin, and P.~J. Sutton.
\newblock {\em Phys. Rev. D}, 46:921, 1992.

\bibitem{supp4}
J.~Kwieci\'nski, A.~D. Martin, and P.~J. Sutton.
\newblock {\em Phys. Lett. B}, 287:254, 1992.

\bibitem{supp5}
J.~Bartels and H.~Lotter.
\newblock {\em Phys. Lett. B}, 309:400, 1993.

\bibitem{supp6}
J.~Bartels.
\newblock {\em J. Phys. G}, 19:1611, 1993.

\bibitem{dane1}
C.~Adloff et~al.
\newblock {H1 Collaboration}.
\newblock {\em Nucl. Phys. B}, 538:3, 1999.

\bibitem{dane2}
{ZEUS Collaboration}.
\newblock {\em Eur. Phys. J. C}, 6:239, 1999.

\bibitem{dane3}
A.~De Roeck.
\newblock {\em Acta Phys. Pol. B}, 29:1343, 1998.

\end{thebibliography}

%
%
%
%
%
%
%
\end{document}